\documentclass[twocolumn,aps,prl]{revtex4}

\usepackage[T1]{fontenc}
\usepackage[latin9]{inputenc}
\usepackage{amsmath}
\usepackage{amssymb}
\usepackage{graphicx}
\usepackage{esint}
\usepackage{color}
\usepackage{mathtools}
\usepackage{float}
\usepackage{placeins}

\definecolor{Green}{RGB}{0,180,0}
\definecolor{Purple}{RGB}{102,0,255}
\definecolor{Blue}{RGB}{51,153,255}
\definecolor{Red}{RGB}{201,010,010}
\definecolor{Ao}{rgb}{0.0, 0.0, 1.0}

\makeatother

\begin{document}

\title{The Peripheral Vortex Biome of Confined Quantum Fluids and Its Influence on Vortex Pair Annihilation}

\author{Chuanzhou Zhu$^1$, Patrick C. Ford$^2$, Mark E. Siemens$^2$, and Mark T. Lusk$^1$}
\affiliation{$^1$Department of Physics, Colorado School of Mines, Golden, CO 80401, USA$^*$\\
	$^2$Department of Physics and Astronomy, University of Denver, Denver, CO 80208, USA}
\email{mlusk@mines.edu, Mark.Siemens@du.edu}

\date{\today}
\begin{abstract}
The self-annihilation of oppositely charged optical vortices in a quantum fluid is hindered by nonlinearity and promoted by radial confinement, resulting in rich life-cycle dynamics of such pairs. The competing effects generate a biome of peripheral vortices that can directly interact with the original pair to produce a sequence of surrogation events. Numerical simulation is used to elucidate the role of the vortex biome as a function of nonlinearity strength and the initial spacing between the engineered vortices. The results apply directly to other nonlinear quantum fluids as well and may be useful in the control of complex condensates in which vortex dynamics produce topologically protected phases. 
	
\end{abstract}
\maketitle

%%%%%%%%%%%%%%%%%%%%%%
\section{Introduction}
%%%%%%%%%%%%%%%%%%%%%%
The ability to anticipate and control vortex dynamics in two-dimensional, nonlinear quantum fluids is scientifically and technologically important to disciplines ranging from quantum turbulence~\cite{Anderson2016} to the generation of non-Abelian anyons~\cite{NonAbelian1,NonAbelian2} and their use in topological quantum computing~\cite{QuantumCompute1,QuantumCompute2,QuantumCompute3,QuantumCompute4,QuantumCompute5}. While the motion of individual vortices can now be predicted~\cite{Zhu2021, Andersen2021,Jasmine}, it is often the nucleation and annihilation of oppositely charged vortex pairs that dominates processes of interest~\cite{superfluid_annihilate}. In Bose-Einstein
condensates (BEC), for instance, such processes have been
observed around a repulsive Gaussian obstacle~\cite{BEC_Obstacle}
or precipitated by stirring~\cite{BEC_stir}, and they have been shown to dominate analogs to quantum turbulence experimentally observed in beams of light~\cite{Alperin2019}.  Vortex-antivortex
lattices have also been predicted for superfluids~\cite{superfluid_lattice}. While these can be studied using Ginzburg-Landau theory~\cite{superconductor_annihilate1,superconductor_annihilate2, superconductor_nucleate1,superconductor_nucleate2}, it is also possible to elucidate few-body vortex nucleation and annihilation using the Gross-Pitaevskii equation~\cite{Gross1961,Pitaevskii1961}. Collective dynamics of many-body vortex systems have been studied in both linear~\cite{Alperin2019} and nonlinear~\cite{many_vortices1,many_vortices2,many_vortices3} quantum fluids, but few-body dynamics of interacting vortices have received less attention. 

Nonlinear optical fluids are also governed by the Gross-Pitaevskii equation~\cite{Agrawal2013} and so serve as controllable, classical analogs to quantum fluids. The optical setting offers
deterministic programming of the initial vortex positions and shapes,
direct readout of evolved states, and room temperature operation.

In the present work, the dynamics and annihilation of vortex pairs are computationally studied in trapped, nonlinear optical quantum fluids. The combination of harmonic trapping and nonlinearity of the medium nonlinearity results in an evolution of nucleation and annihilation events in  which the the original vortex pair interacts with vortices nucleated at the trap boundary. The analysis is framed within the setting of nonlinear optical fluids, but the results apply to any two-dimensional fluid governed by the Gross-Pitaevskii equation. In the limiting regime of linear and weakly nonlinear media, the results are consistent with previous investigations~\cite{optical_fluid2,Indebetouw,Jasmine, Bao}. Beyond the perturbative setting though, our numerical simulations reveal qualitatively new behavior in which the role of the boundary vortices is elucidated. A wide range of fluid nonlinearities and initial vortex separations are used to generate a phase diagram delineating settings in which the primary vortices either do or do not annihilate as a function of initial vortex separations and medium nonlinearity. 

%%%%%%%%%%%%%%%%%%%%%%
\section{Setting}
%%%%%%%%%%%%%%%%%%%%%% 

Consider a linearly polarized, monochromatic electromagnetic wave with an electric field of the form
\begin{equation}
	\vec E(\vec r_\perp, z, t) = \psi(\vec r_\perp, z) e^{ik_0(z-c_0 t)}\vec e_x .
\end{equation}
The free-space wavenumber is $k_0$, $c_0$ is the speed of light in vacuum, $\vec e_x$ is the unit vector orientation of the polarization, $\psi$ is a scalar measure of the electric field, the transverse position vector is $\vec r_\perp$, and the beam axial coordinate, $z$, plays the role of time. In fact, we will refer to changes associated with travel along this axis as time evolution.  

For media with a third-order susceptibility and beam profiles for which the paraxial approximation~\cite{Lax1975} is valid, Maxwell's equations then imply that the evolution of the scalar field, $\psi$, is well-approximated by the Gross-Pitaevskii equation:
\begin{equation}
	i\, \partial_z \psi = \left( -\frac{1}{2 k_0 n_0}\nabla_\perp^2  - k_0 \Delta n  - 
	k_0 n_2 |\psi|^2 \right) \psi.  \label{eq:nonlinearParaxial}
\end{equation}
Here $n_0$ and $n_2$ are the index of refraction and nonlinear refractive indices of the medium, and $\Delta n$ is an external change in the index of refraction,
\begin{equation}\label{trap}
	\Delta n = -\frac{1}{2}\gamma r^2,
\end{equation} 
that is generated by either a patterned medium or an additional laser beam. The dielectric profile curvature, $\gamma$, has units of inverse length squared and must be sufficiently small, so that $\gamma<<k_0^2$. %Dielectric_Trapping_of_Vortices_031721.nb%
This condition allows additional terms involving the dieletric profile to be safely neglected. Optical nonlinearity gives rise to a photon-photon interaction of strength $-k_0 n_2$ that can be either attractive ($n_2>0$) or repulsive ($n_2<0$). Within this setting, the propagation of light is formally analogous to the dynamics of two-dimensional quantum fluids at the mean-field level~\cite{Carusotto2014, Ozawa2019}. The introduction of a characteristic length of $1/(k_0 n_0)$ and characteristic field intensity of $I_0=\epsilon_0 c_0 E_0^2/2$ allows the position variables and scalar electric field to be expressed non-dimensionally. The resulting evolution equation is
\begin{equation} \label{eq:GPE}
	i\, \partial_z \psi = \left( -\frac{1}{2}\nabla_\perp^2  + \frac{\omega^2}{2}r^2 + \beta|\psi|^2 \right) \psi.
\end{equation}
Here $r^{2}=x^{2}+y^{2}$, and
\begin{equation}\label{beta}
	\beta = -\frac{\tilde n_2 I_0}{n_0}  \quad {\rm and} \quad \omega = \sqrt{\frac{\gamma}{k_0^2 n_0^3}}
\end{equation}
are the non-dimensional strength of the medium nonlinearity and the non-dimensional trap strength, respectively, and $\tilde n_2$ is the nonlinear index of refraction in terms of intensity, $\Delta n = \tilde n_2 I_0$.

This theoretical framework can be applied to experiments on both optical fluids~\cite{optical_fluid1,optical_fluid2,Jasmine} and Bose-Einstein condensates~\cite{Cornell,Spielman,Polkinghorne,GPELab2}, with $\beta$ on the order of $1000$ achievable in both settings.

The 2D Gross-Pitaevskii equation is numerically solved using an unconditionally stable time splitting pseudo-spectral method~\cite{GPELab1,GPELab2} by discretizing the spatial domain into a $1024*1024$ grid. The domain size is $\pm 20$ in both $x$ and $y$ directions, and the evolution step size is prescribed as $10^{-3}$.

The GPE of Eq.~(\ref{eq:GPE}) is numerically solved for initial states for which a Gaussian profile is implanted with two diametrically-opposed, oppositely-charged vortices that are offset from the center of the fluid by $x_{0}$:
\begin{equation}\label{eq:initial_state}
\psi_{0}\left(x,y\right)=\frac{N}{\sqrt{\pi}}e^{-\frac{1}{2}\left(x^{2}+y^{2}\right)}v_{+}\left(x,y\right)v_{-}\left(x,y\right).
\end{equation}
Here $N$ is a normalization factor, and the initial profiles of the
vortices are 
\begin{equation}\label{eq:initial_vortices}
v_{\pm}\left(x,y\right)=\sqrt{1-\frac{l^{2}}{l^{2}+\left(x\mp x_{0}\right){}^{2}+y^{2}}}\frac{\left(x\mp x_{0}\right)\pm iy}{\sqrt{\left(x\mp x_{0}\right)^{2}+y^{2}}}.
\end{equation}
The healing length, $l$, is chosen to make the vortex structures as
stable as possible (Eq. 3.17 in \cite{Barenghi}):
\begin{equation}
l=\frac{1}{\sqrt{\beta}\psi_{bg}}.\label{eq:healing_length}
\end{equation}
Here $\psi_{bg}$ is the background fluid amplitude at the vortex initial position, $\{\pm x_{0},0\}$, located at the shoulder of a Gaussian profile. It is given by
\begin{equation}
\psi_{bg}=\frac{1}{\sqrt{\pi}}e^{-\frac{1}{2}x_{0}^{2}}.
\label{eq:psi_B_initial}
\end{equation}
Note that the repulsive interaction ($\beta>0$) in a nonlinear fluid tends to compress vortex footprints which results in periodic fluctuations in the core size of evolving vortices.  Such core-size fluctuations cannot be completely eliminated, since the fluid amplitude is not evenly distributed around a vortex moving inside the trapped fluid.
However, enforcement of the relationship between $l$ and $\beta$ in Eq.~(\ref{eq:healing_length}) best minimizes the influence of this effect~\cite{Zhu2021}.

In the limit of a linear fluid $(\beta=0)$, the core approaches a linear profile: %profiles are also linear: 
\begin{equation}
v_{\pm}\left(x,y\right)=\left(x\mp x_{0}\right)\pm iy.\label{eq:initial_vortices_LC}
\end{equation}
%

%%%%%%%%%%%%%%%%%%%%%%%%%%%%%%%%%%%%
\section{Linear Fluids}
%%%%%%%%%%%%%%%%%%%%%%%%%%%%%%%%%%%%  

%%%%%%%%%%
% FIGURE 1
\begin{figure}
	\includegraphics[width=0.5\linewidth]{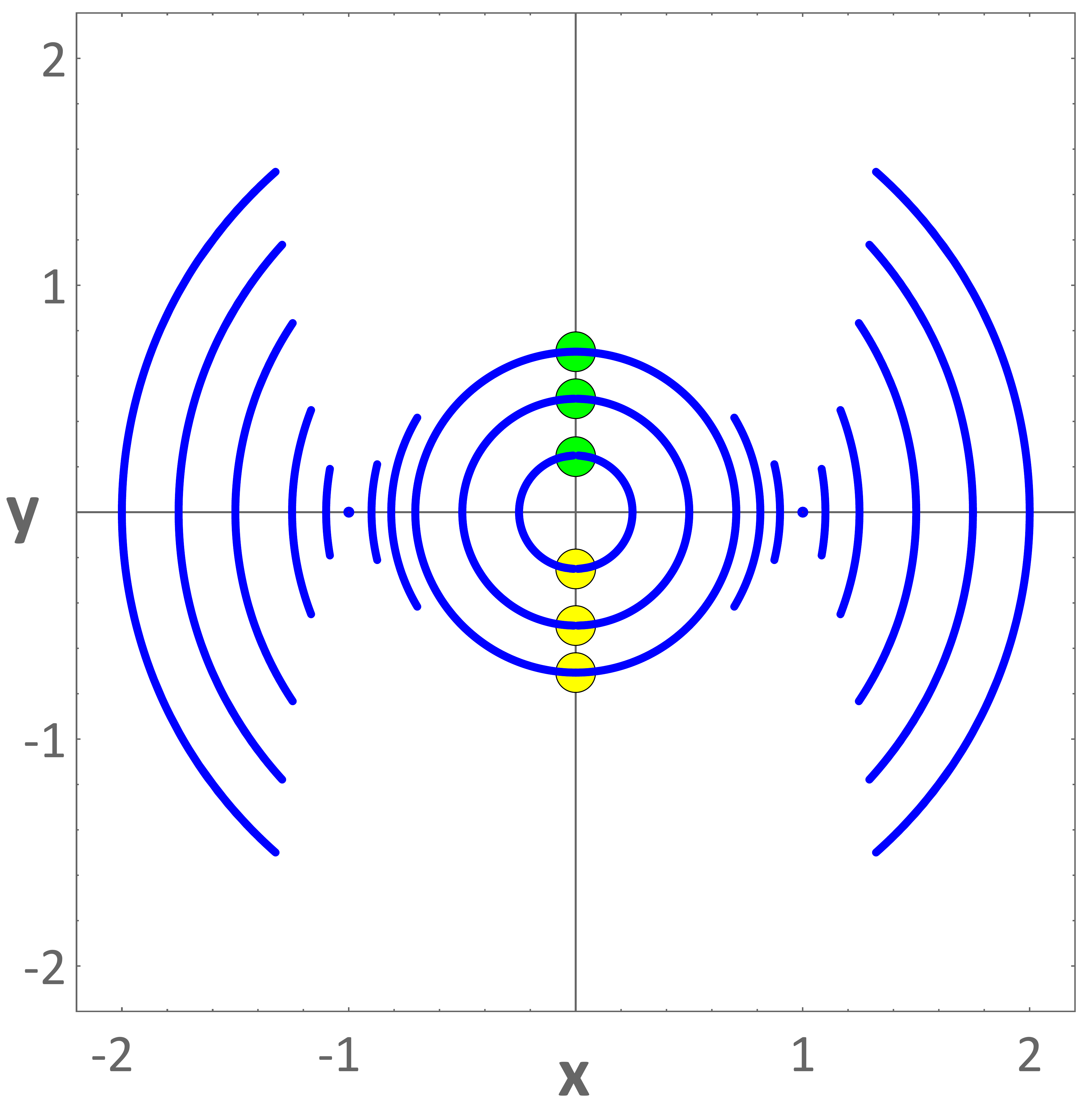}
	\caption{Analytical trajectories for linear, trapped fluid, generated with Eqs. (\ref{eq:xv_trap}) and (\ref{eq:yv_trap}), over a range of initial vortex separations. For initial separations less than $1/\sqrt{2}$, the vortices annihilate at the bottom of their arcs (yellow disks), but after an intervening interval, re-nucleate at this same spot. Then the two vortices reverse their course and repeat the annihilation/nucleation at the top (green disks). For initial separations greater than $1/\sqrt{2}$, each vortex moves back and forth on an arc that does not intersect with its partner.}
	\label{Fig_Linear}
\end{figure}
%%%%%%%%%%

First consider trapped vortex dynamics for linear media ($\beta=0$), where we will see that no peripheral vortices are spawned, and analytical trajectories of the primary pair have been previously derived~\cite{Bao}. This gives a baseline for subsequent consideration of nonlinear effects. The initial state is given by Eqs. (\ref{eq:initial_state}) and (\ref{eq:initial_vortices_LC}), and the trap strength in
Eq.~(\ref{eq:GPE}) is chosen as $\omega=1$ to match the size of the mode. The evolving  state is obtained in closed form with the vortex trajectories then extracted: 
\begin{align}
x_{v}\left(t\right) & =\pm\sqrt{x_{0}^{2}-\bigl(y_{v}\left(t\right)\big)^{2}},\label{eq:xv_trap}\\
y_{v}\left(t\right) & =\left(x_{0}-\frac{1}{x_{0}}\right)\sin\left(t\right).\label{eq:yv_trap}
\end{align}
These analytical trajectories for linear fluid are plotted in Fig.~\ref{Fig_Linear}, over a range of initial vortex separations. For initial separations of $x_{0}<1/\sqrt{2}$, the two vortices cyclically annihilate and nucleate with an intervening interval during which no vortices exist. 
The timing of annihilation and nucleation can be obtained by substituting Eq.~(\ref{eq:yv_trap}) into Eq.~(\ref{eq:xv_trap}) and solving the equation $x_{v}\left(t\right)=0$. The intervening interval is the time difference between the first and second solutions of this equation. 
For initial separations $x_{0}>1/\sqrt{2}$, the two vortices cyclically trace out only a fraction of these semicircular arcs and never meet. 
%This is shown in Fig. \ref{Fig_Linear}, and the distinction in regimes is identified with a thick red line at the bottom of Fig. \ref{Fig_Phase_Diagram}. \mlnote{We need to find a better way of relating each case with the phase diagram. Simplify referring to a figure that appears at the beginning or end of the paper is pedagogically awful.}

%%%%%%%%%%%%%%%%%%%%%%%%%%%%%%%%%%%%
\section{Nonlinear Fluids}
%%%%%%%%%%%%%%%%%%%%%%%%%%%%%%%%%%%%  

Nonlinear quantum fluids do not exhibit such simple cyclic behavior. Some headway can be made using a perturbative analysis provided the nonlinearity is sufficiently weak~\cite{Bao}, but a complete picture of the dynamics emerges only with a numerical study over a wider range of nonlinearity strength. The evolution of vortices in trapped, nonlinear quantum fluids exhibits a new feature, a seething annulus of vortex nucleation and annihilation at the periphery of the central portion of the trap. For the most part, the vortices in this rich biome live and die far away from the trap center and do not meaningfully affect the primary vortex pair. However, select members of this ecosystem play an essential role in the life cycle of the central pair. 

\begin{figure}
	\includegraphics[width=0.99\linewidth]{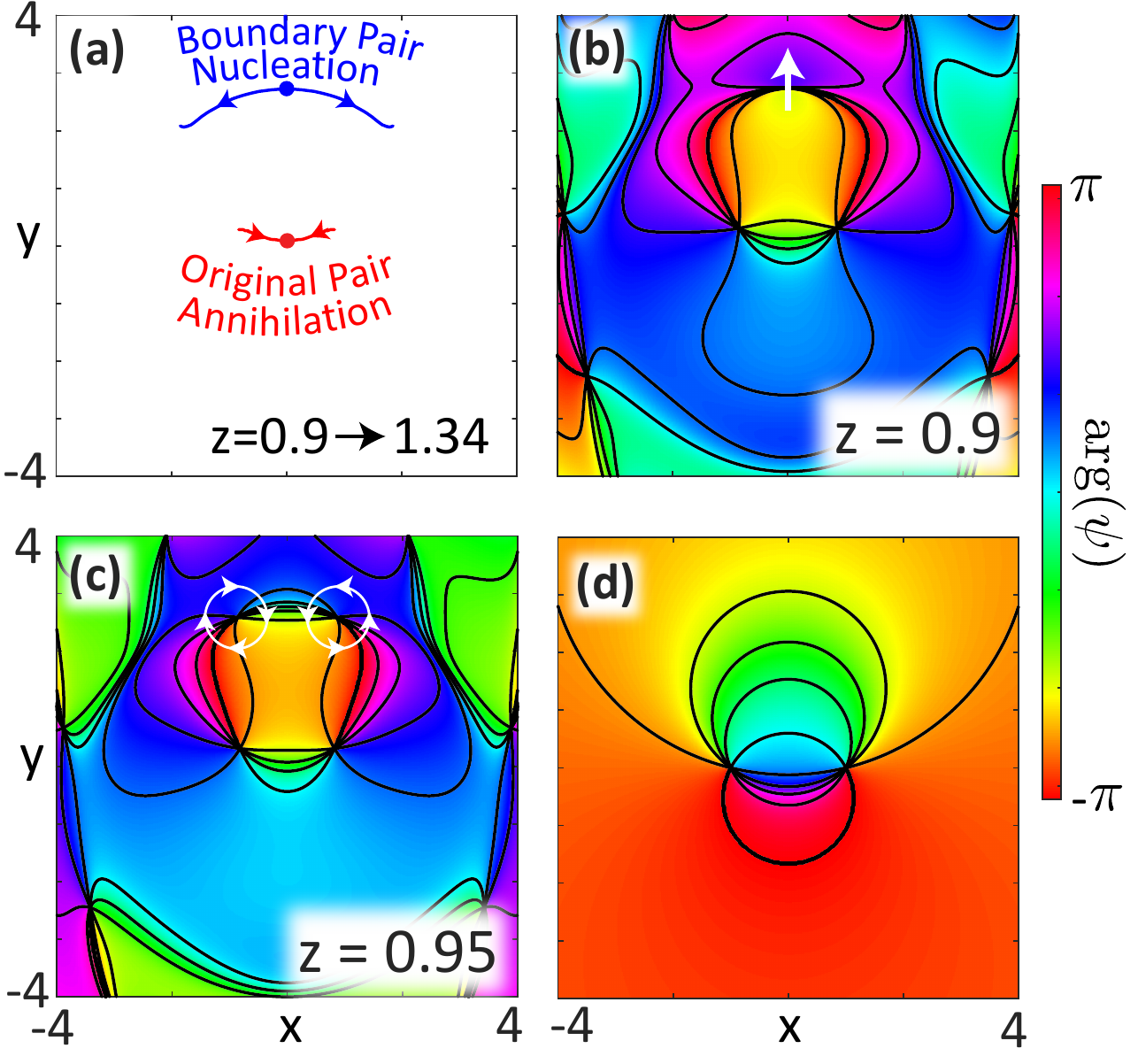}
	\caption{Vortex evolution in a trapped fluid for $\beta=10$ and $x_{0}=1$. (a) The trajectories of two original vortices (red) and two boundary vortices (blue) during the time period from the nucleation of the boundary vortex pair at $z=0.9$ (blue dot) to the annihilation of the original vortex pair at $z=1.34$ (red dot); (b)(c) The fluid phases just before and after the nucleation of the boundary vortex pair, where the black contours are phase lines and the white arrows denote the	directions of local fluid velocity--i.e. the negative of the phase gradient; (d) For comparison, the fluid phase for a trapped, linear fluid with $\beta=0$ and $x_{0}=1$ at $z=0.9$.}
	\label{Fig_B10}
\end{figure}

The same initial conditions are applied, but now with the vortex healing length set by Eq. 8 and trap frequency set to
\begin{equation}
	\omega=\sqrt{1+\frac{\beta}{2\pi}}.
	\label{eq:trap_frequency}
\end{equation}
This minimizes the fluid and vortex core expansion induced by the repulsion from the finite $\beta$. The combination of nonlinearity and beam confinement produces a peripheral biome of vortices.  Fig.~\ref{Fig_B10} provides a particularly clear example of this.  Although there are multiple boundary pairs, here we focus on the way in which nucleation occurs for the boundary pair closest to the center. The time slices of the fluid phases in panels (b) and (c) capture this nucleation event. The white arrow in panel (b) denotes a strong local, upward fluid velocity in a necked region, identifiable because the black phase contour lines are very dense. This calls to mind the pulling of an oar through water that results in oppositely spinning whirlpools on either end, and such settings are known to generate vortex pairs~\cite{narrow_velocity}. The resulting vortices are shown in panel (c).

While this explains how vortices form in regions with sufficiently steep phase gradients, it does not explain how such gradients arise. For that, it is useful to consider that, for a linear trapped fluid, all the phase contours are circular, as shown in panel (d). Because of this, boundary vortex pairs will never nucleate in a linear trapped fluid. The addition of nonlinearity, though, induces phase inhomogeneities within the trap setting; even a slight deformation of nonlinear fluid in a trap may result in substantial fluid flow in the boundary annulus. In fact, we find that \emph{any} nonzero value of nonlinearity, $\beta$, will result in the emergence of a boundary vortex biome.

%%%%%%%%%%
% FIGURE 2
% Plot_Evol_5.m in the subfolder x0_1_beta_10 in the folder 211102_Run_Opposite_Charge_trap
%%%%%%%%%%

%%%%%%%%%%%%%%%%%%%%%%%%%%%%%%%%%%%%
\subsection{Vortex Annihilation for Nonlinear Fluids}
%%%%%%%%%%%%%%%%%%%%%%%%%%%%%%%%%%%%   

%%%%%%%%%%
% FIGURE 3
\begin{figure}
	\includegraphics[width=0.99\linewidth]{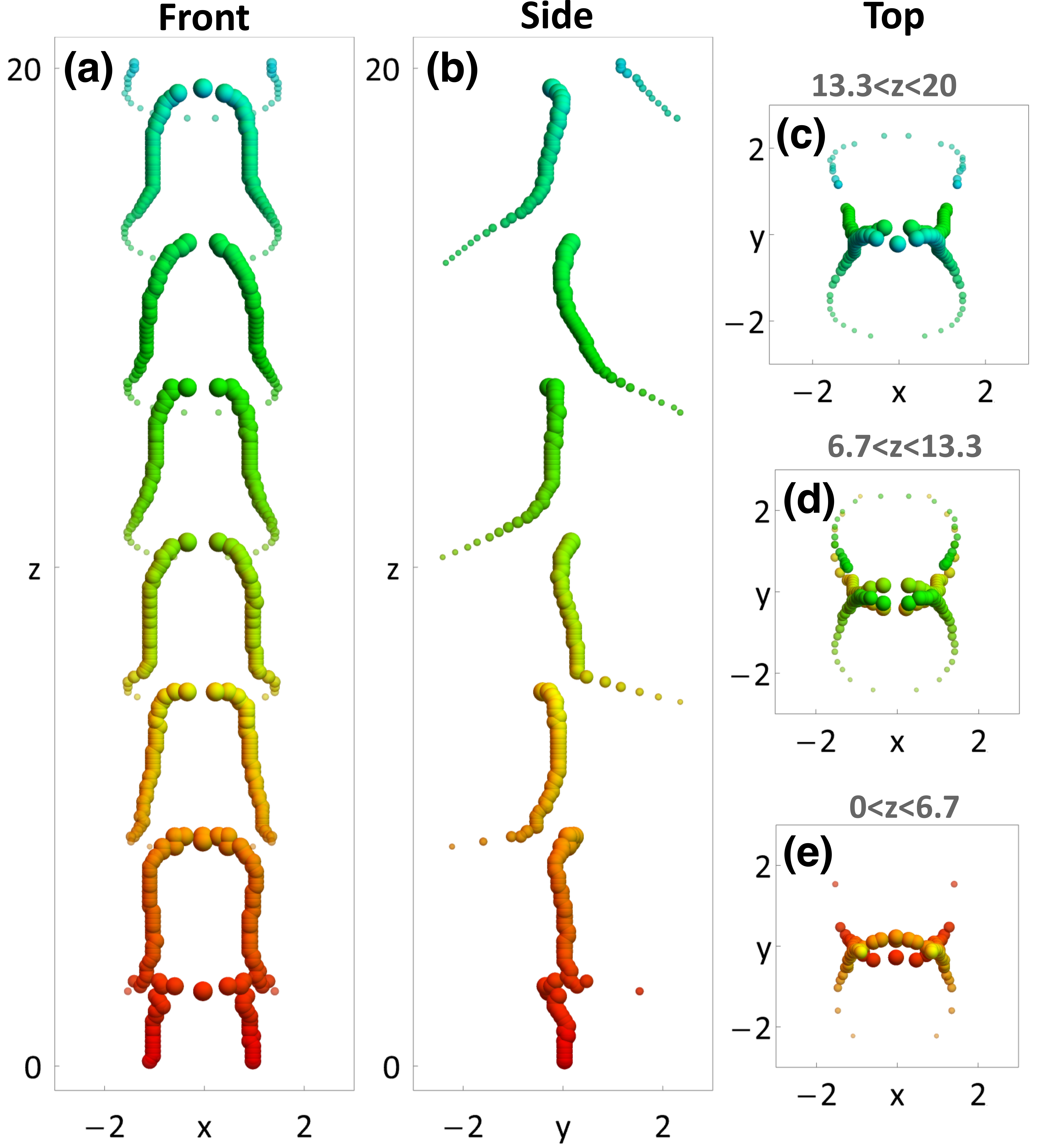}
	\caption{Three-dimensional plot of vortex positions in a trapped laser beam propagating along the $z$ direction for $x_0=1$ and $\beta=1$. Panel (a) is the front view; panel (b) is the right-side view; and panel (c-e) are the top-down views during three different $z$ periods. The color range from red to cyan corresponds to the evolution from $z=0\rightarrow20$. Sphere size is inversely proportional to distance from the center. Evolution runs vertically from bottom to top.}
	\label{Fig_B1}
\end{figure}
% Data in the subfolder x0_1_t_20 in the subfolder x0_1 in the folder 211102_Run_Opposite_Charge_trap
% File: Manuscript_Dot_Plots.nb
%%%%%%%%%%

%%%%%%%%%%
% FIGURE 6
\begin{figure}
	\includegraphics[width=0.99\linewidth]{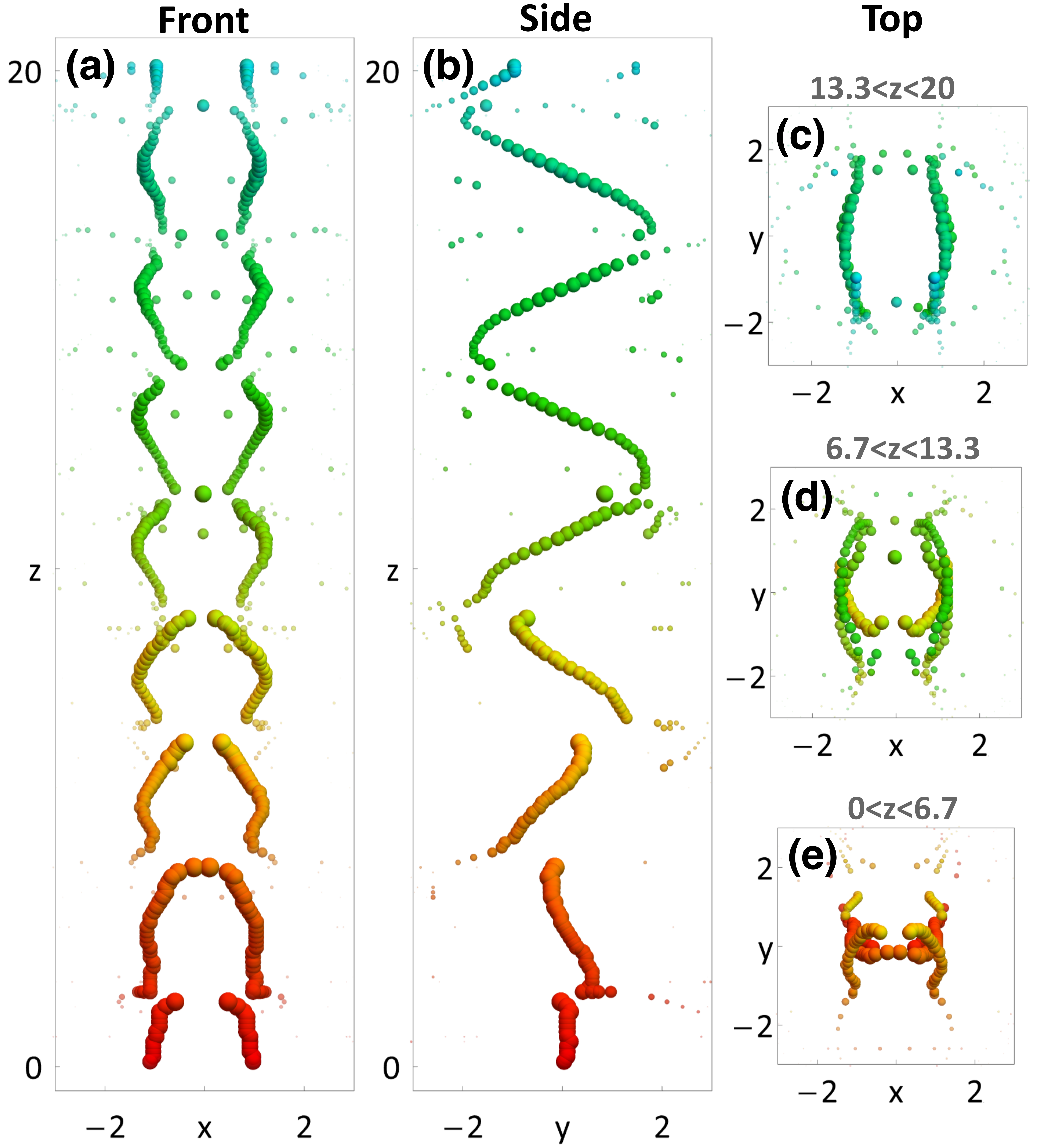}
	\caption{Three-dimensional plot of vortex positions in a trapped laser beam propagating along the $z$ direction for $x_0=1$ and $\beta=6$. Panel (a) is the front view; panel (b) is the right-side view; and panel (c-e) are the top-down views during three different $z$ periods. The color range from red to cyan corresponds to the evolution from $z=0\rightarrow20$. Sphere size is inversely proportional to distance from the center. Evolution runs vertically from bottom to top.}
	\label{Fig_B6}
\end{figure}
% Data in the subfolder x0_1_t_20 in the subfolder x0_1 in the folder 211102_Run_Opposite_Charge_trap
% File: Manuscript_Dot_Plots.nb
%%%%%%%%%%%

The annulus of boundary vortices may indirectly influence the trajectories of the primary vortex pair, but a surprisingly intimate relationship actually exists between the two types of vortices. Vortex dynamics in weakly nonlinear, trapped quantum fluids are shown in Figs.~\ref{Fig_B1} and \ref{Fig_B6} for an initial separation of $x_{0}=1$ and nonlinearity strengths of $\beta=1$ and $\beta=6$, respectively. Spheres represent vortex position, with diameter inversely proportional to distance from the trap center. This makes the many vortex pairs at the periphery essentially invisible. For this initial separation, the vortex positions are stationary in the linear fluid, as shown in Eqs.~(\ref{eq:xv_trap}) and~(\ref{eq:yv_trap}), but 
they annihilate in even weakly nonlinear fluids. Fig.~\ref{Fig_B1} shows that the primary vortex pair annihilates, but that a boundary vortex pair is nucleated even before this occurs. These boundary vortices rapidly move into the central region of the trap and take up positions near those initially occupied by the primary pair. The process then repeats, with the surrogate pair annihilating even as a new boundary pair nucleates. 

A new time scale becomes apparent as the nonlinearity is increased to $\beta=6$, as shown in Fig.~ \ref{Fig_B6}. Surrogate pairs periodically nucleate at the boundary, but the y-range of the central pair increases with each cycle. This occurs, albeit more slowly, for the case shown in Fig. \ref{Fig_B1} as well. An increase in nonlinearity strength causes the primary vortices and their surrogates to carry out circuits that evolve outwards, cycle by cycle, resulting in trajectories that are closer to the boundary region of the trap. 

\begin{figure}
	\includegraphics[width=0.99\linewidth]{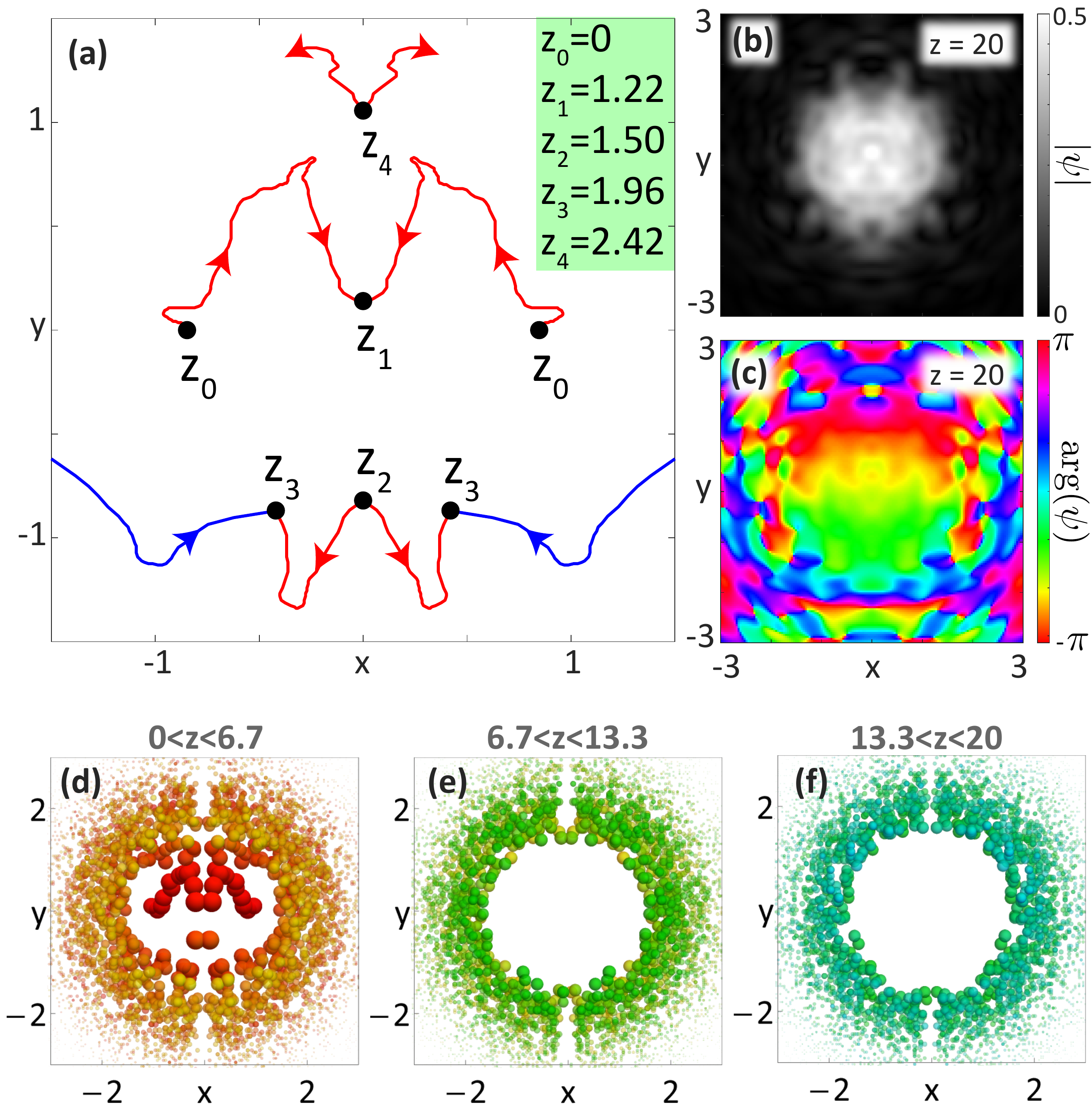}
	\caption{Vortex evolution for $\beta=120$ and $x_{0}=0.85$. (a) Trajectories of the primary vortices for the time period from $z=0$ to $z=2.6$, where the primary vortex pair annihilates at $z_{1}$, nucleates at $z_{2}$, annihilates with boundary vortices at $z_{3}$, and nucleates again at $z_{4}$. (b,c) Fluid amplitude and phase at $z=20$.  (d-f) Plots for the positions of all vortices during three evolution periods, $0<z<6.7$, $6.7<z<13.3$, and $13.3<z<20$, where the color range from red to cyan corresponds to the evolution from $z=0\rightarrow20$. Sphere size is inversely proportional to distance from the trap center. Evolution runs vertically from bottom to top.}
	\label{Fig_B120}
\end{figure}

An even larger nonlinearity can be used to determine the ultimate fate of the expanding surrogation cycles. Nascent boundary pair surrogates interact more and more strongly with other boundary vortices, and the distinction is lost between surrogate boundary pairs and other boundary vortices. An example of this is shown in Fig.~\ref{Fig_B120}, for $\beta=120$ and $x_0=0.85$. Eventually, the primary vortex pair self-annihilates, but the nascent surrogates are waylaid en route to the trap center and are themselves annihilated by another pair of boundary vortices. A steady state is then reached in which all vortices have been swept out of the central region. The system settles into a structure in which the boundary vortex biome is divided into left and right regimes with a sparsely populated central channel. Fig.~\ref{Fig_B120} shows this limiting type of evolution, and the final state structure, for $\beta=120$ and $x_0=0.85$. The vertical channel of low vortex density that coalesces is the result of very rapid vortex speeds associated with nucleation and annihilation, showing that the sequence identified in panel (a) continues.

Analogous dynamics and trends characterize the vortex motion for smaller values of initial vortex separation, $x_0$, although the vortex biome is thinner and the rate of assimilation is slower. This fundamental dynamical character actually hold for all values of $x_0$ provided the nonlinearity is sufficiently small. However, there is an entire region of the $\{x_0, \beta\}$ parameter space for which the central vortices never annihilate, as will now be elucidated.

% Plot_Vortex_Position_Tilt.m in the subfolder beta120 in the subfolder x0_0.85 in the folder 211102_Run_Opposite_Charge_trap
% File: Manuscript_Dot_Plots.nb
%%%%%%%%%%

%%%%%%%%%%%%%%%%%%%%%%%%%%%%%%%%%%%%
\subsection{No Annihilation for Highly Nonlinear Fluids}
%%%%%%%%%%%%%%%%%%%%%%%%%%%%%%%%%%%%   

\begin{figure}
	\includegraphics[width=0.96\linewidth]{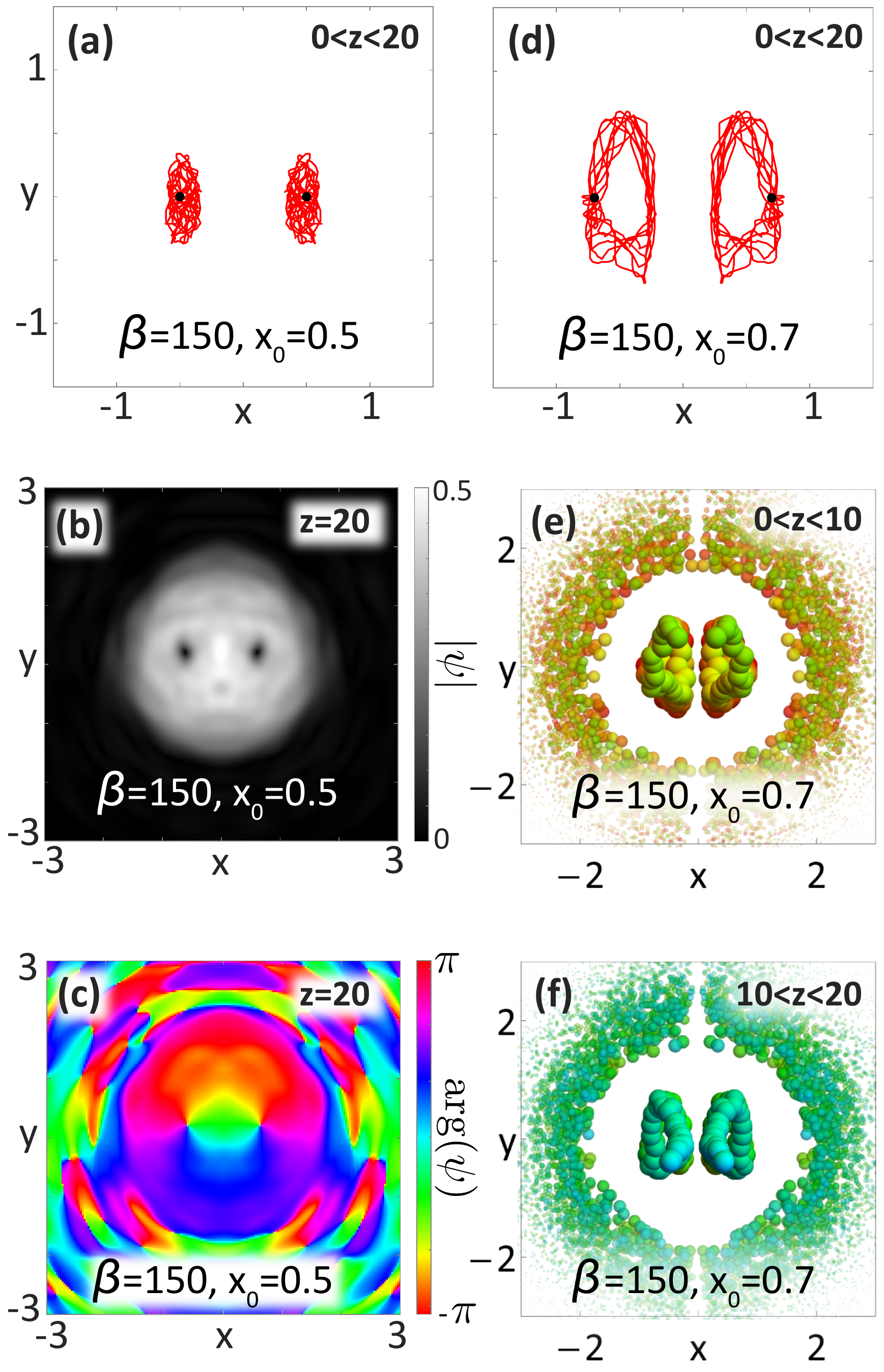}
	\caption{Vortex evolution for $\beta=150$. (a) The trajectories ($0<z<20$) of the two central vortices with initial offset, $x_{0}=0.5$. (b,c) Fluid amplitude  and phase at $z=20$ corresponding to the end of the trajectory in panel (a). (d) The trajectories ($0<z<20$) of the two central vortices for $x_{0}=0.7$. (e,f) Plots of the positions of all vortices during sequential intervals, $0<z<10$ and $10<z<20$, for $x_{0}=0.7$ with colors corresponding the "time": red ($z=0$) to cyan ($z=20$). Sphere size is inversely proportional to distance from the trap center. Evolution runs vertically from bottom to top.}
	\label{Fig_B150}
\end{figure}

For sufficiently large values of $\beta$, the influence of nonlinearity overwhelms boundary-vortex effects and annihilation of the central vortex pair is no longer possible.  The trajectories of Figure~\ref{Fig_B150} are associated with this regime. This is shown in Fig. \ref{Fig_B150} with $\beta=150$. An initial offset of $x_0=0.5$ causes the central vortices to simply wobble slightly, and the system reaches a  steady state with the associated fluid amplitude and phase plotted in panels (b) and (c), respectively. The positions and circular cross-sections of the vortices are essentially unchanged, and a thick, well-separated biome shield wall exists at the periphery. As the initial separation is increased, the wobble of the central vortices becomes more pronounced. This is shown in panels (d)-(f). 

%%%%%%%%%%
% FIGURE 8  
% Plot_Vortex_Position_Tilt.m in the subfolders x0_0.5_beta150_t0to20 and x0_0.7_beta150_t0to20 in the folder 211102_Run_Opposite_Charge_trap
% File: Manuscript_Dot_Plots.nb
%%%%%%%%%%

\section{Phase Diagram}

\begin{figure}
	\includegraphics[width=0.99\linewidth]{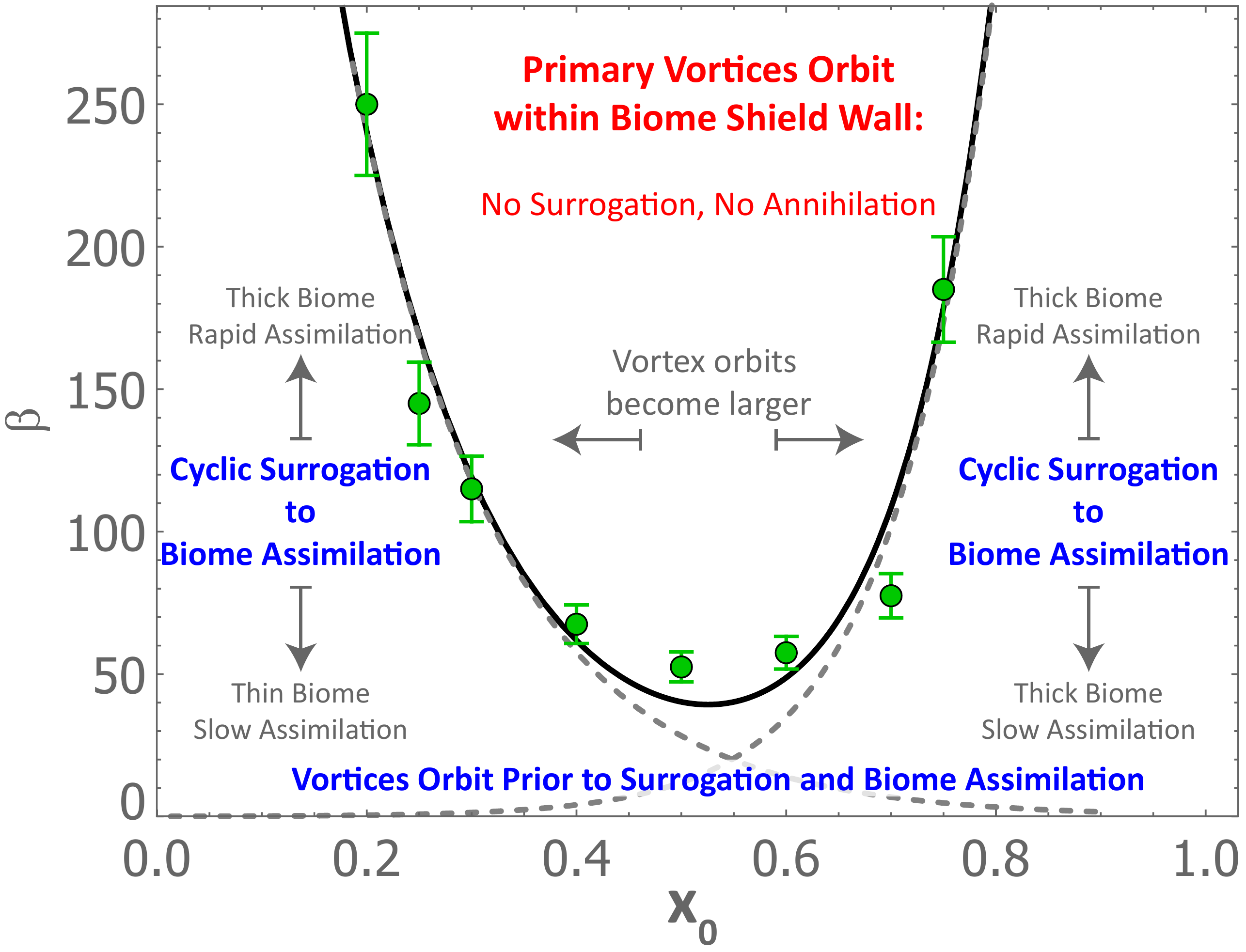}
	\caption{Dynamical phase diagram for the annihilation of the original vortex pair in nonlinear quantum fluid, in the parameter space of initial vortex separation, $x_{0}$, versus fluid nonlinearity, $\beta$. The trap frequency is set to $\omega=\sqrt{1+\beta/\left(2\pi\right)}$. The black solid curve is the fit of Eq.~(\ref{eq:ftrap}) for the phase boundary. The two original vortices annihilate in region below the phase boundary, while they do not annihilate in the region above the phase boundary. Data points (green dots) are the result of numerical simulation with error bars indicating $\pm 5\%$ of the critical $\beta$ for each point.}
	\label{Fig_Phase_Diagram}
\end{figure}

We are now in a position to map out the behavior of a pair of oppositely charged vortices as a function of their initial separation and the strength of fluid nonlinearity. An extensive set of simulations, over a grid of values of $x_{0}$ and $\beta$, was used to produce the phase diagram shown in Fig.~\ref{Fig_Phase_Diagram}. A parabolic-shaped phase boundary (black solid curve) separates the parameter space into two regions for which original vortices either do or do not annihilate. Green dots correspond to phase boundary points for which no annihilation occurs out to $z=20$. Their locations do not change when the simulation time is doubled. A grid of neighbouring values of $\beta$ was tested to determine that the critical values are accurate to within $\pm 5\%$ of the critical $\beta$. 
	
The cupped-shape phase boundary was then fitted to the sum of an exponential decay and an exponential rise (grey dashed curves) using a nonlinear optimization routine:
\begin{equation}
f_{\rm trap}(x_{0})=Ae^{-bx_{0}}+Ce^{dx_{0}}.
\label{eq:ftrap}
\end{equation}
An optimal fit was found for $A=997.0$, $b=7.125$, $C=0.05667$, and $d=10.70$, resulting in the black curve in Fig.~\ref{Fig_Phase_Diagram}.
%Phase_Diagram_Nonlinear_Vortex_Annihilation.nb

%%%%%%%%%%
% FIGURE 7
% Phase_Diagram_Nonlinear_Vortex_Annihilation.nb
% phase_diagrams.m in the folder 211102_Run_Opposite_Charge_trap
%%%%%%%%%%

%%%%%%%%%%%%%%%%%%%%%%%%%%%%%%%%%%%%
\section{Discussion}
%%%%%%%%%%%%%%%%%%%%%%%%%%%%%%%%%%%%   

%The beam is guided within a medium that has a harmonic dielectric profile to trap these vortices.  The beam is assumed to satisfy the paraxial approximation, and the propagation axis is interpreted as time. The phase diagrams were generated by numerically solving the two-dimensional Gross-Pitaevskii equation for a grid of initial vortex separations and fluid nonlinearities. While framed within an optical setting, the phase diagrams are equally valid for any two-dimensional quantum fluid whose behavior can be approximately described by the Gross-Pitaevskii equation, such as Bose-Einstein condensates.

The construction of a phase diagram allows the observed trends to be put together into a coherent story.  Fluid nonlinearity reduces vortex healing length and effectively reduces the influence of the vortices on one another, thus making them less inclined to annihilate. The fluid is trapped, though, and boundary effects compete against this by promoting vortex annihilation. The boundary influence is weak for vortices with small separation, so a very large nonlinear effect is required to prevent their annihilation. The nonlinearity required for this is less when the vortex separation is larger since they have a lower mutual attraction~\cite{Andersen2021}. If their initial separation is increased sufficiently, though, the boundary effect becomes prominent, driving the vortex pair towards annihilation. To prevent this, the nonlinearity must be increased. It is this competition that explains the parabolic shape of the phase boundary.  

In all regions of the phase diagram, the presence of a biome of boundary vortices plays a role in the dynamics of the central vortex pair. Below the phase boundary, the original pair will self-attract and annihilate as happens for linear fluids. This is immediately followed, though, by a surrogation process in which a boundary pair rushes in to take the place of the original vortices. The process then repeats while, on slower time scale, the surrogate pairs become more widely separated and are more strongly influenced by other biome vortices. At some point, an annihilation in the trap center is not replaced because the nascent surrogates are waylaid and annihilated by other biome vortices. The result is a central region that is swept clear of all vortices and surrounded by a stable, clearly delineated shield biome.

Above the phase boundary, there is a balance between the competing forces of nonlinear myopia and external trap pressure. The latter pushes each vortex into a wobbling orbit about its initial position while the former amounts to a repulsive force that keeps them from combining. Since it is the global change in field generated by their annihilation that causes nascent surrogate vortices to rush in, the biome does not produce any such candidates and a stable shield wall is maintained.    

However, for large initial separation of the primary pair and relatively low nonlinearity,  self-annihilation of the primary pair is always followed by its replacement with a surrogate boundary pair that was forming and moving towards the center even as the primary pair was undergoing self-annihilation. Higher values of nonlinearity cause the nascent boundary pair to annihilate with other boundary pairs, and the result is a central trap region swept clear of any vortices at all. Figure \ref{Fig_Phase_Diagram} has been annotated to identify key trends such as the rate of assimilation, biome structure, and size of vortex orbits.

For highly nonlinear media, it is possible to freeze the primary vortex pair in place within a clean central region. This offers a unique way to prepare a pair of stable, oppositely-charged vortices with circular cross-sections, something that is simply not possible in the absence of both nonlinearity and trapping.

In a Bose-Einstein condensate, the value of $\beta$ is determined by the product of the atom number and the two-atom interaction strength. In a typical experiment for a $^{87}\rm{Rb}$ condensate with the atom number $10^5$~\cite{Cornell} and the natural atomic interaction strength $7.79\times10^{-12}\rm{Hz} \cdot \rm{cm}^{3}$~\cite{Spielman}, the strength of the nonlinearity can be estimated
to be on the order of $\beta=100$. The value of $\beta$ can be further adjusted through Feshbach resonance. In fact, $\beta=500$ has been regarded as a reasonable estimation by previous works on vortices in Bose-Einstein condensates~\cite{Polkinghorne,GPELab2}.

In  optical fluids, it has been shown that a positive (self-defocusing) third-order nonlinearity can be induced by propagating a laser field within a hot atomic rubidium vapor~\cite{Fontaine}. The strength of the nonlinearity was reported to be on the order of $\beta = 10^{-6}$, yet still yield nonlinear effects such that the system displays a Boguliobov dispersion relation in which low-wave number features propagate like phonons, and high-wavenumber features propagate like interacting particles. Higher values of $\beta$ have been shown ~\cite{doi:10.1126/science.aae0330} in indium tin oxide using a modestly-powered pulsed laser ($2.5\times10^{15} W\cdot m^{-2}$, $200 fs$) and producing a nonlinearity on the order of $\beta = 25$. It is conceivable that even higher nonlinearities can be achieved in other materials ~\cite{alma991041409138402766} and using higher intensity beams. It is important to reiterate that vortex biome effects should be present for any value of $\beta$ greater than zero.

While the vortex biome and annihilation dynamics described here have not been experimentally demonstrated, work in precision control and readout of  nonlinear quantum fluids suggests that such measurements could be achievable in the next few years. Light propagating nonlinearly in a medium is a promising system, but those measurements have so far been done in a bulk medium with no confinement -- quite different from the harmonic traps considered in this paper that led to the vortex biome. Laser pre-patterning of the dielectric constant in the material is one possible route toward providing the external trap. Another challenge is taking measurements at different propagation steps through a nonlinear medium \cite{vocke2015experimental,barsi2009imaging}. Alternatively, atomic BEC could also host the same vortex dynamics with a straightforward means of trapping, but precision vortex preparation and non-perturbative readout remain significant challenges.

We are grateful to the W.M. Keck Foundation and the National Science Foundation (DMR 1553905) for supporting this research.

%\bibliographystyle{prsty} 	
%\bibliography{NonlinearAnnihilation}

\end{document}